\documentclass[aps,amssymb,floatfix,prd,amsmath,preprintnumbers]{revtex4}
\usepackage{epstopdf}
\usepackage{capt-of}
\usepackage{graphicx}  
\usepackage{dcolumn}   
\usepackage{bm}
\begin{document}

\title{\bf Accelerating Dark Energy Cosmological Model in Two Fluid with Hybrid Scale Factor }

\author{B. Mishra, P. K. Sahoo, Pratik P. Ray\\
Department of Mathematics,\\
Birla Institute of Technology and Science-Pilani, Hyderabad Campus,\\
Hyderabad-500078, India\\
bivudutta@yahoo.com,  sahoomaku@rediffmail.com, pratik.chika9876@gmail.com}

\affiliation{ }

\begin{abstract}
\begin{center}
\textbf{Abstract}
\end{center}
In this paper, we have investigated the anisotropic behaviour of the accelerating universe in  Bianchi V space time in the frame work of General Relativity (GR). The matter field we have considered is of two non interacting fluids i.e. the usual string fluid and dark energy (DE) fluid. In order to represent the pressure anisotropy, the skewness parameters are introduced along three different spatial directions. To achieve a physically realistic solutions to the field equations, we have considered a scale factor, known as hybrid scale factor, which is generated by a time varying deceleration parameter. This simulates a cosmic transition from early deceleration to late time acceleration. It is observed that the string fluid dominates the universe at early deceleration phase but does not affect nature of cosmic dynamics substantially at late phase where as, the DE fluid dominates the universe in present time, which is in accordance with the observations results. Hence, we analysed here the role of two fluids in the transitional phases of universe with respect to time which depicts the reason behind the cosmic expansion and DE. The role of DE with variable equation of state parameter (EoS), skewness parameters also discussed along with physical and geometrical properties.
\end{abstract}

\maketitle
\section{Introduction}
There has been a continuous trial to extend our understanding regarding the dark driven late time cosmic expansion. The direct observations and evidences by the High-z supernovae search teams \citep{riess1, perl1} revealed that we live in an accelerated expanding universe. The present state of accelerated expansion of the universe was first indicated by Solheim \cite{Soleh66}. Observing luminosity of several cluster of galaxies, it was indicated that the model giving the best fit to the data are with a non vanishing cosmological constant and negative deceleration parameter. Further, the claim has gained significant attention after the more recent discoveries of Supernova Ia \cite{perl2, perl3, riess2, riess1}, CMBR anisotropy \citep{cald1,huang}, Large scale structure \citep{seljak, tegmark}, Baryon Acoustic Oscillation (BAO) \citep{eisenstein}, Weak lensing \citep{jain}. From these observations, it has been observed that the understanding on the present universe is flat geometrically. However, the most revolutionary and counter intuitive result from these observations is that $4.6\%$ of the total energy budget is non-relativistic or baryonic matter and $24\%$ non-baryonic matter, called the dark matter. The remaining $71.4\%$  is an unknown form of exotic cosmic fluid with negative pressure, called DE. Despite the gravitational attraction of matter, DE provide a strong negative force leading to an anti gravity effect that drives the acceleration \citep{astier1, copeland, cald2, silve}. Many theoretical and observational investigations have been carried out in order to know the real nature of DE. The main problem in the context is the fact that DE does not interact with baryonic matter and hence there is no way to detect this mysterious component. Also in GR, DE corresponds to an isotropic fluid with negative pressure and constant energy density .\\

From different observational results, cosmologists argued that cosmological constant $\Lambda$ is the most acceptable candidate for the DE. The reason behind this argument is that during cosmic evolution, it has the constant energy density with negative pressure. However, this concept of DE can be invoked and had a checked history it often suffers "fine tuning" and "cosmic coincidence" puzzle \citep{copeland}. Hence,the different forms of dynamically changing DE with an effective EoS parameter $ \omega= \frac{p}{\rho},$ where $p$ is the pressure and  $\rho$ is energy density of matter has been proposed in many literature. However, at present time, the exact value of DE equation of state parameter is still unknown. Our lack of knowledge allows us to suggest different candidates for DE. Besides the consideration of cosmological constant with $\omega=-1$ ($\Lambda$CDM model), canonical scalar field models such as quintessence ($-\dfrac{2}{3}\leq \omega \leq- \dfrac{1}{3}$) \citep{ratra, sahni}, phantom field($\omega<-1$) \citep{cald3}, k-essence \citep{armen1, armen2}, tychons \cite{sen} , quintom \cite{feng} and the interacting DE candidates chaplygin gas \cite{ bento, kamen}, holographic models \cite{wang, setare1, setare2}, brane world \cite{daff, li} have been proposed as varieties of possible solutions of DE problem. In these models, a parametrized form of DE density is considered and basing upon the requirements the parametrized form turned to get viable models.\\

Recently current cosmological data from Supernovae 1a \citep{astier2, riess2}, cosmic microwave background radiation \citep{eisenstein, mac} and large scale structures \citep{koma} have checked the possibility of $\omega\ll -1$, where as the DE crossing the phantom divide line ($\omega\ll -1$) is mildly favored. Some alternate limits obtained from observational results \citep{knop, tegmark} are $-1.67 < \omega < -0.62 $ and $-1.33 < \omega <-0.79$ respectively. As of now, we only know that DE is non clustering and spatially homogeneous; although its effect was small in early time, it dominates the present universe. Besides the issue of late time cosmic acceleration, the observed anisotropy in temperature power spectrum is another point of discussion in recent times. Though the universe is isotropic at large scale according to the prediction of $\Lambda$ CDM model, it is expected to have small scale anisotropies in the universe supported by the observations of CMB radiation data from WMAP and Planck. These data shows some non trivial topology of the large scale geometry of the universe with an asymmetric expansion \citep{campa1, campa2, campa3, gruppo}. From the theoretical arguments and experimental data, the existence of an anisotropic universe with anisotropic pressure has been confirmed, and which ultimately approaches to isotropic universe. Therefore, it is worthy to study the models of the universe with anisotropic background in the presence of DE.\\

Recently, Mishra et al. \citep{mishra1}  have constructed anisotropic DE cosmological models in Bianchi  type V space time in GR by considering  different matter field. Pacif and Mishra \citep{pacif} constructed the anisotropic DE cosmological model by constraining the Hubble parameter. Again, Mishra and Tripathy \citep{mishra2} suggested a cosmological model, where they have considered a variable  deceleration parameter with a hybrid scale factor. This consideration simulates a cosmic transition from early deceleration to late time acceleration. Most of the models with constant deceleration parameter have been studied by considering perfect fluid or ordinary matter in the universe. But neither of those matter is enough to describe the dynamics of an accelerating universe. This motivates the researchers to consider the model of the accelerating universe filled with mixed fluid, i.e, some exotic matter such as the DE along with some usual matter or fluid such as cosmic string fluid. String cosmological model have generated a lot of research interests in recent times because of their roles in describing different interesting phenomena. During the phase transition in the early universe with spontaneous symmetry breaking, string arise as a random network of line-like defects. After the initiation of the general relativistic theory of string cloud cosmology by Letelier \citep{letelier} and Stachel \citep{stachel}, many researchers have investigated different aspects of the string cosmological models in the framework of GR showing the dominance of cosmic string fluid in early universe.\\

In the context of two fluid matter such as dark fluid matter along with usual ordinary matter (Baryonic matter), a number of literature has motivated the researchers to investigate different models in the back drop of GR relativity with different Bianchi forms. Akarsu and Kilinc \citep{akarsu1, akarsu2} have investigated the DE cosmological models in Bianchi I and Bianchi III space time with constant deceleration parameter. Yadav et al. \citep{yadav1} has presented locally rotationally symmetric Bianchi V universe with DE characterized by variable EoS parameter assuming constant deceleration parameter. Theoretical models of interacting and non interacting DE have been discussed widely in the literature \citep{amir1, amir2, pradhan1, pradhan2, shey, zimda}. Tripathy et al. \citep{tripa1} investigated Bianchi V DE model with two non interacting fluids as matter field and found the pressure anisotropy which continues along with the cosmic expansion. Also they showed those models were mostly dominated by phantom behaviour. In this present work, we have constructed anisotropic DE model at the back ground of Bianchi V space time in GR and investigated the dynamics of the model considering two different non interacting fluids contributing to the matter field. The first one is usual cosmic string fluid and the second one is due to the DE. The outline of the paper is as follows; In section 2, we have set up the mathematical formulation of the problem along with its corresponding physical parameters. In section 3, the significance of hybrid scale factor has been analysed which mimic a cosmic transition. The physical and kinematical properties of the model has been discussed in section 4. In section 5, the results and summaries are discussed.
 
\section{Mathematical Formalism of the model}
We have considered a spatially homogeneous and anisotropic Bianchi V space time in the form
\begin{equation} \label{eq:1}
ds^{2}= dt^{2}-[A(t)]^{2}dx^{2}-e^{2\alpha x}([B(t)]^{2}dy^{2}+[C(t)]^{2}dz^{2})
\end{equation}
where, the metric potentials $A$, $B$ and $C$ are functions of $t$ only. $\alpha$ is non zero arbitrary constant. The energy momentum tensor for the given environment with two non interacting fluids is,

\begin{equation}
T_{ij}=T_{ij}^{(cs)}+T_{ij}^{(de)} \label{eq:2}
\end{equation}
where $T_{ij}^{(cs)}$ and $T_{ij}^{(de)}$ respectively denote the contribution to energy momentum tensor from one dimensional cosmic string and DE. For cosmic string,the energy momentum tensor can be defined as 

\begin{equation}\label{eq:3} 
T^{(cs)}_{ij} = (\rho + p) u_{i} u_{j}- pg_{ij}+ \lambda x_{i}x_{j}
\end{equation}
where $u_{i}u_{j}=x_{i}x_{j}=1$ (along x- direction). In a co moving coordinate system $u^i$ is the four velocity vector of the fluid. The pressure $p$ for this fluid is taken to be isotropic, $\rho$ as proper energy density for a cloud of strings and $\lambda$ as string tension density. The energy momentum tensor for DE can be described as

\begin{eqnarray} \label{eq:4}
T^{(de)}_{ij}  = diag[-\rho_{de}, p_{de_x},p_{de_y},p_{de_z}]\\ \nonumber
			   = diag[-1, \omega_{de_x},\omega_{de_y},\omega_{de_z}]\rho_{de}\\ \nonumber
			   =diag[-1, \omega_{de}+\delta,\omega_{de}+\gamma,\omega_{de}+\eta]\rho_{de} ,
\end{eqnarray}

where $\omega_{de}$ is the equation of state parameter (EoS)for DE and $\rho_{de}$ is the DE density. The skewness parameters, the deviations
from $\omega_{de}$ on $x$, $y$ and $z$ axes are respectively represented as $\delta$, $\gamma$ and $\eta$ . Now, the field equations for a two fluid i.e the cosmic string and DE with  Bianchi type V space-time in the frame work of GR are obtained as  

\begin{equation}\label{eq:5}
\frac{\ddot{B}}{B}+\frac{\ddot{C}}{C}+\frac{\dot{B}\dot{C}}{BC}-\frac{\alpha^{2}}{A^{2}}=-p-\lambda-(\omega_{de}+\delta)\rho_{de}  
\end{equation} 

\begin{equation}\label{eq:6}
\frac{\ddot{A}}{A}+\frac{\ddot{C}}{C}+\frac{\dot{A}\dot{C}}{AC}-\frac{\alpha^{2}}{A^{2}}=-p-(\omega_{de}+\gamma)\rho_{de} 
\end{equation}

\begin{equation}\label{eq:7}
\frac{\ddot{A}}{A}+\frac{\ddot{B}}{B}+\frac{\dot{A}\dot{B}}{AB}-\frac{\alpha^{2}}{A^{2}}=-p-(\omega_{de}+\eta)\rho_{de}
\end{equation}

\begin{equation}\label{eq:8}
\frac{\dot{A}\dot{B}}{AB}+\frac{\dot{B}\dot{C}}{BC}+\frac{\dot{C}\dot{A}}{CA}-\frac{3\alpha^{2}}{A^{2}}=\rho+\rho_{de} 
\end{equation}

\begin{equation} \label{eq:9}
2\frac{\dot {A}}{A}-\frac{\dot{B}}{B}-\frac{\dot{C}}{C}=0
\end{equation}

where an over dot is represented as the derivatives with respect to time $t$. Eqn. \eqref{eq:9} yields $A^2=k BC$. Suitably the integration constant absorbed into the metric potential $B$ or $C$ to yield $A^2=BC$. In order to study the models, some physical parameters can be defined as : the cosmic spatial volume and the scalar expansion $\theta$ can be obtained respectively as $V=R^3=ABC$ and $\theta=3H$, where $H=\frac{\dot{R}}{R}=\frac{1}{3}(H_x+H_y+H_z)$.  $H_x$, $H_y$ and $H_z$ are directional Hubble parameter in the direction of $x$, $y$ and $z$ respectively. The shear scalar $\sigma^2=\frac{1}{2}(H_x^2+H_y^2+H_z^2-\frac{\theta^2}{3})$ and the deceleration parameter can be calculated from $ q=-\frac{R\ddot{R}}{\dot{R}^2} $. In order to get an anisotropic relation between the scale factors of different directions, we assume $B=C^m, m\neq1$. Accordingly the scale factors in different directions can be expressed as , $A=R, B=R^{\frac{2m}{m+1}}, C=R^{\frac{2}{m+1}}$. As the behaviour of the model would be studied through the behaviour of Hubble parameter, therefore the directional Hubble parameters can be calculated as, $H_x=H$, $H_y=\big(\frac{2m}{m+1}\big)H$ and $H_z=\big(\frac{2}{m+1}\big)H$. The anisotropic parameter $\mathcal{A}$ can be obtained as

\begin{equation} \label{eq:10}
{\mathcal A}=\frac{1}{3}\sum \biggl(\frac{\vartriangle H_i}{H}\biggr)^2
\end{equation}

The energy conservation equation for the anisotropic fluid, $T^{ij}_{;j}=0$ can be defined as 

\begin{equation} \label{eq:11}
\dot{\rho}+3(p+\rho)H+\lambda H_x+\dot{\rho_{de}}+3\rho_{de}(\omega_{de}+1)H+\rho_{de}(\delta H_x+\gamma H_y+\eta H_z)=0
\end{equation}

From eqn. \eqref{eq:11}, the energy conservation equation for string fluid, $T^{ij(cs)}_{;j}=0$ and DE fluid $T^{ij(de)}_{;j}=0$ respectively obtained as:

\begin{equation} \label{eq:12}
\dot{\rho}+3(p+\rho)H+\lambda H_x=0
\end{equation}
and
\begin{equation} \label{eq:13}
\dot{\rho_{de}}+3\rho_{de}(\omega_{de}+1)H+\rho_{de}(\delta H_x+\gamma H_y+\eta H_z)=0
\end{equation}

We assume $\rho=\xi \lambda$ and $p=\omega \rho $ in \eqref{eq:12}. The constants, $\xi$ and $\omega$ are state parameter variables which can be positive and negative and  $\omega \in[-1,1]$. A positive $\xi$ value indicates the presence of one dimensional string in cosmic fluid and a negative $\xi$ value shows the strips of the string phase of the universe. Hence, we obtain
\begin{equation} \label{eq:14}
\rho=(ABC)^{-(1+\omega)}A^{-\frac{1}{\xi}}\rho_0
\end{equation}
where $\rho_0$ is the integration constant or rest energy density. Subsequently, pressure and string density can be obtained as:

\begin{eqnarray} 
p=\frac{\omega \rho_0 A^{-\frac{1}{\xi}}}{(ABC)^{1+\omega}};\label{eq:15}\\
\lambda=\frac{\rho_0 A^{-\frac{1}{\xi}}}{\xi (ABC)^{1+\omega}} \label{eq:16}
\end{eqnarray}

Now, eqn. \eqref{eq:8} and eqn. \eqref{eq:14} yields,

\begin{equation}\label{eq:17}
\rho_{de}= \frac{\dot{A}\dot{B}}{AB}+\frac{\dot{B}\dot{C}}{BC}+\frac{\dot{C}\dot{A}}{CA}-\frac{3\alpha^{2}}{A^{2}}-\frac{\rho_0 A^{-\frac{1}{\xi}}}{(ABC)^{1+\omega}}
\end{equation}

We can split \eqref{eq:13} into two parts; one corresponds to the deviation of EoS parameter and other is deviation free part as:

\begin{equation} \label{eq:18}
\dot{\rho_{de}}+3\rho_{de}(\omega_{de}+1)H=0 
\Rightarrow \rho_{de}=\rho_{de_0} R^{-3(\omega_{de}+1)}
\end{equation}

\begin{equation} \label{eq:19}
\rho_{de}(\delta H_x+\gamma H_y+\eta H_z)=\rho_{de} \left(\delta +\gamma \left(\frac{2m}{m+1}\right)+\eta \left(\frac{2}{m+1}\right)\right)\frac{\dot{R}}{R}=0  
\end{equation}
According to the eqn. \eqref{eq:18}, the DE density behaviour $(\rho_{de})$ is diagnosed by the deviation free part of EoS parameter of DE , where $\rho_{de_0}$ indicates DE density at present epoch. The second equation deals with the anisotropic part i.e. the deviations from the total DE pressure along different spatial directions.\\

Algebraic manipulation of equations \eqref{eq:5} ,\eqref{eq:6}, \eqref{eq:7} and
using $\lambda$ from eqn. \eqref{eq:16}, the skewness parameters take the form,
\begin{eqnarray}
\gamma= \frac{1}{3\rho_{de}}\biggl[\frac{(m+5)}{(m+1)} \chi(m) F(R)+\frac{\rho_0}{\xi} G(R)\biggr] \label{eq:20}\\
\eta=-\frac{1}{3\rho_{de}}\biggl[\frac{(5m+1)}{(m+1)} \chi(m) F(R)-\frac{\rho_0}{\xi} G(R)\biggr] \label{eq:21}\\
\delta=-\frac{2}{3\rho_{de}}\biggl[\frac{(m-1)}{(m+1)} \chi(m) F(R)-\frac{\rho_0}{\xi} G(R) \biggr] , \label{eq:22}
\end{eqnarray}
where  $\chi (m)= \dfrac{m-1}{m+1}$, represents the amount of difference from isotropic behaviour of the model.   $F(R)=\biggl( \dfrac{\ddot{R}}{R}+ \dfrac{2 \dot{R}^{2}}{R^{2}} \biggr)$ and $G(R)=\dfrac{1}{R^{3(1+ \omega+\frac{1}{3\xi})}}.$\\
To get the model isotropic, $\chi(m)$ vanishes and consequently,
the skewness parameters will retain only the string terms, because the anisotropic part vanished. Subsequently, the DE density $\rho_{de}$ and the EoS parameter $\omega_{de}$ can be  obtained as,

\begin{align} \label{eq:23}
\rho_{de} =   \dfrac{\dot{R}^2}{2R^2}\left[ \phi(m) \right]- 3 \dfrac{\alpha ^2}{R^2}- \rho_{0} G(R) 
\end{align}

\begin{align} \label{eq:24}
- \omega_{de} \rho_{de} & = \dfrac{4\ddot{R}}{3 R} \left[ \phi(m) \right]  +  \dfrac{2\dot{R}^2}{3R^2} \left[ \phi(m) \right] + \rho_{0} G(R) \left[ \dfrac{1}{3 \xi} + \omega \right] - \dfrac{\alpha ^{2}}{R^2}, 
\end{align}

where, $\phi(m)= \dfrac{m^{2} + 4 m + 1}{(m+1)^2}$.

The functional $F(R)$ and the DE density $\rho_{de}$ completely controlled the behaviour of skewness parameters and the functional $\chi (m)$ depends upon the exponent $m$. EoS parameter for DE is decided by the anisotropic nature of the model through this
exponent $m$, the time dependence of mean Hubble parameter $H$, the hybrid scale factor $R$ and the rest energy density $\rho_{0}$. It is interesting to note that, the DE density and the DE EoS parameter depend on the usual cosmic string fluid too. In the absence of usual cosmic string fluid, the above eqns.  \eqref{eq:23} and \eqref{eq:24} reduce to the earlier expressions of Ref. \cite{mishra1}.

\section{Hybrid scale factor and Cosmic transit}

Based on the recent outcomes on the present universe, the appropriate choice for the scale factor are the de Sitter solution and the power law expansion. The behaviour of both power law and exponential law of the scale factor lead to a constant deceleration parameter. However, the time dependence of the directional scale factor would be decided by specific choices of scale factors. In the present work, we have considered the specific scale factor, the hybrid scale factor which at late time results into a constant deceleration parameter. The hybrid scale factor has two factors in the form, $R=e^{at}t^{b}$, where $a$ and $b$ are positive constants. In this combined form of the scale factor one form is on exponential expansion and the other is on power law expansion. The cosmic dynamics is dominated by the power law $(t^{b})$ in the early phase, whereas it is dominated by the exponential factor $(e^{at})$ at late phase. However, eventually, the hybrid scale factor pulls itself away towards later phase of evolution in a more dominant fashion which may mimic the accelerated expansion of universe. When $a=0,$ the scale factor reduces to the power law and for $b=0,$ the exponential law recovered. So, a cosmic transition from early deceleration to late time acceleration can be obtained by a hybrid scale factor.\\

For this model, the Hubble parameter and the directional Hubble parameter can be obtained respectively as $H = a+\dfrac{b}{t}$ and $H_{x} = a+\dfrac{b}{t}, H_{y}=\dfrac{2m}{m+1}\left(a+\dfrac{b}{t}\right) $ and $Hz = \dfrac{2}{m+1}\left(a+\dfrac{b}{t}\right)$ \citep{akarsu3, kumar2, pradhan3}. Tripathy \citep{tripa2} has considered a general form of such  parameter in the form $H = a+ \dfrac{b}{t^{n}}$ , $n$ being a constant.\\

The deceleration parameter $q = - \dfrac{\ddot{R}}{R^2}$ describes cosmic dynamics concerning the late time acceleration. A positive deceleration parameter indicates a decelerating universe, negative value describes a universe with acceleration expansion and a null value predicts a universe expanding with constant rate. Since most of observational data favour an accelerating universe, models with late time cosmic dynamics have got much focus in recent times. In this present work, we are interested in the kinematics of late time accelerating
universe while the deceleration parameter is assumed to vary slowly or becoming a constant with cosmic time. According to type Ia Supernovae observations, the deceleration parameter for the accelerating universe in the present time is $q = -0.81 \pm 0.14$ \cite{rapet}. Type Ia Supernovae data in combination with BAO and CMB observations declare the deceleration parameter as $q = -0.53 ^{+0.17}_{-0.13}$ \cite{giost}. Eventually, the deceleration parameter becomes $q =   -1+ \dfrac{b}{(at+b)^{2}}$ . At an early phase of cosmic evolution when $t \longrightarrow 0,$ $ q \simeq -1 + \dfrac{1}{b}$ and at late phase of cosmic evolution with $t \longrightarrow \infty , q \simeq -1 $. To find a transient universe with early deceleration and late acceleration, we constrain the parameter $b$ to be in the range $0 < b < 1$ so that at early time $q$ can be positive whereas at late time $q$ assumes a negative value in conformity with the recent observational data.  We can infer from deceleration parameter evaluated here that, the cosmic transit occurs at a time $t = -\dfrac{b}{a}\pm \dfrac{\sqrt{b}}{a}$. The negativity of the second term leads to a concept of negative time which may be unphysical in the context of Big Bang cosmology and therefore, the cosmic transit may have occurred at time $t =\dfrac{\sqrt{b}-b}{a}$, which again restricts $b$ in the same range $0 < b < 1$. The variation of deceleration parameter vs. time for the representing value of the constant $a=0.1$ and $ b=0.3$ has been depicted in Fig. 1 which also shows that the transition occurs at $t=2.47$.\\

\begin{figure}[h!]

\minipage{0.40\textwidth}
\includegraphics[width=75mm]{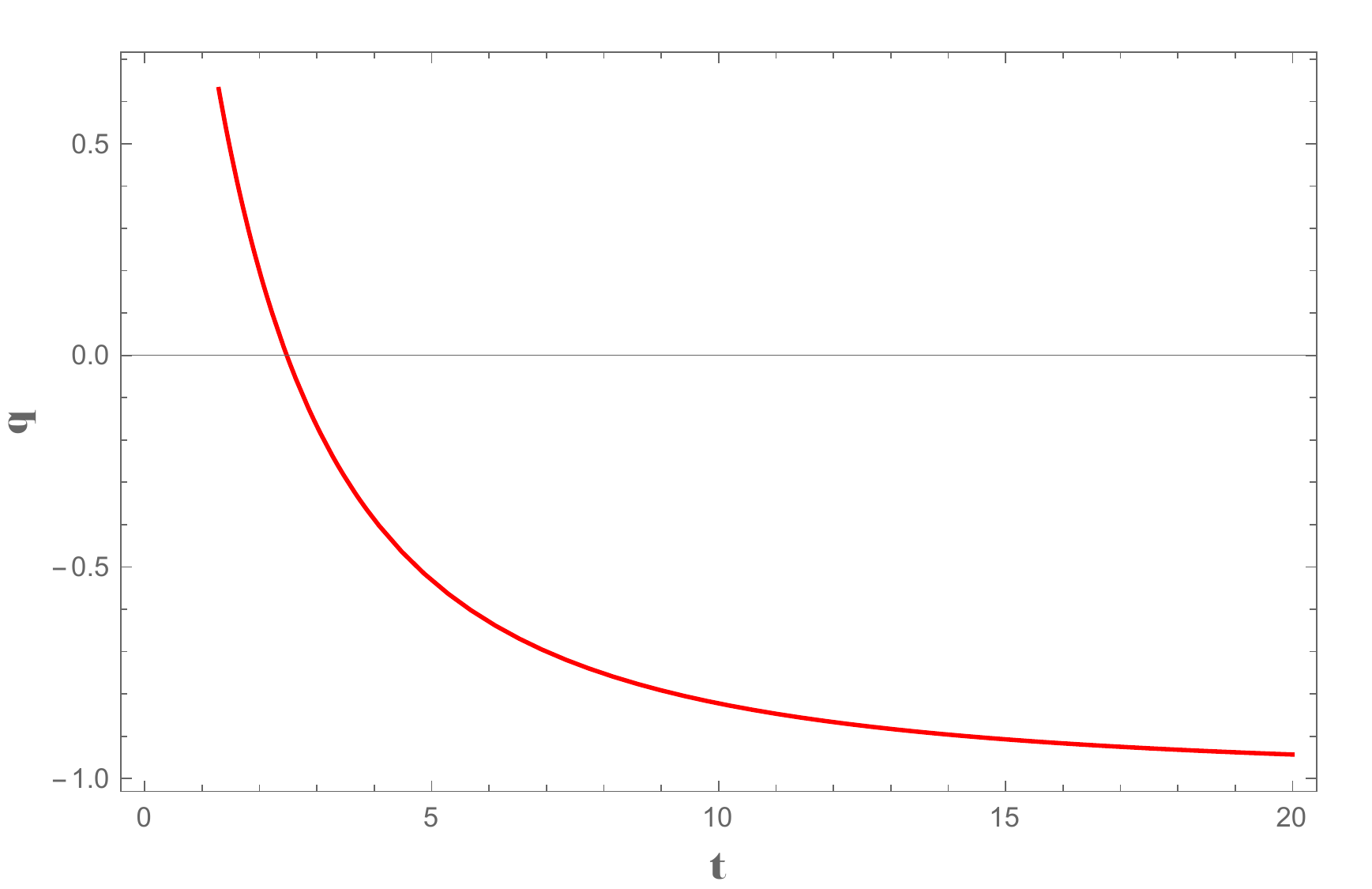}
\caption{Variation of deceleration parameter $q$  vs. time with a=0.1 and b=0.3}
\endminipage
\end{figure}

With this consideration of hybrid scale factor the directional scale factors can be expressed as $ A=R=e^{at}t^b; B=R^{\frac{2m}{m+1}}=(e^{at}t^b)^{\frac{2m}{m+1}};$ and $ C=R^{\frac{m}{m+1}}=(e^{at}t^b)^{\frac{2}{m+1}}$. Subsequently, eqns. \eqref{eq:20} - \eqref{eq:23} can be recasted as, 

\begin{eqnarray}
\gamma &=& \frac{1}{3\rho_{de}}\biggl[\frac{(m+5)}{(m+1)} \chi(m) F(t)+\frac{\rho_0}{\xi} \dfrac{1}{R^{3(1+ \omega-\frac{1}{3\xi})}}\biggr] \label{eq:25}\\
\eta &=&-\frac{1}{3\rho_{de}}\biggl[\frac{(5m+1)}{(m+1)} \chi(m) F(t)-\frac{\rho_0}{\xi} \dfrac{1}{R^{3(1+ \omega-\frac{1}{3\xi})}}\biggr] \label{eq:26}\\
\delta &=& -\frac{2}{3\rho_{de}}\biggl[\frac{(m-1)}{(m+1)} \chi(m) F(t)-\frac{\rho_0}{\xi} \dfrac{1}{R^{3(1+ \omega-\frac{1}{3\xi})}}\biggr] , \label{eq:27}
\end{eqnarray}

where $F(t)=\frac{3a^2t^2+6abt+3b^2-b}{t^2}$. The DE density and EoS parameter can be expressed as,

\begin{align} \label{eq:28}
\rho_{de} =  \frac{2(m^2+4m+1)(a+\frac{b}{t})^2-
 3 \alpha^2(e^{at}t^b)^{-2}(m+1)^2}{(m+1)^2} -  \rho_0(e^{at}t^b)^{\frac{1}{\xi}-3(1+\omega)}
\end{align}

\begin{align} \label{eq:29}
\omega_{de}& = \frac{-4(m^2+4m+1)(a^2 t^2+2abt+b(b-1))}{6(m^2+4m+1)(at+b)^2-9 \alpha^2(m+1)^2 e^{-2at-2b+2} -3 \rho_{0} (m+1)^{2} t^{2} (e^{at}t^{b})^{\frac{1}{\xi}-3(1+\omega)} } \nonumber
\\  &\nonumber -
\frac{2(m^3+5m^2+5m+1)(at+b)^2}{6(m^2+4m+1) (m+1)(at+b)^2-9 \alpha^2(m+1)^3 e^{-2at-2b+2}-3 \rho_{0} (m+1)^{3} t^{2} (e^{at}t^{b})^{\frac{1}{\xi}-3(1+\omega)} } \nonumber \\&
- \frac{[(\frac{1}{3\xi}+\omega)\rho_0(e^{at}t^b)^{-\frac{1}{\xi}-3(1+\omega)}-\alpha^2(e^{at}t^b)^{-2}](m+1)^2}{2(m^2+4m+1)(a+\dfrac{b}{t})^2 - 3 \alpha^{2} (e^{at}t^{b})^{-2} (m+1)^2 - \rho_{0} (m+1)^{2}(e^{at}t^{b})^{\frac{1}{\xi}-3(1+\omega)} }
\end{align}

The density parameter of the cosmic string fluid $\Omega_{cs}$  and the density parameter of DE fluid $\Omega_{de}$ can be derived respectively as 

\begin{align} \label{eq:30}
\Omega_{cs}= \dfrac{\rho_{0}}{3} \dfrac{(e^{at}t^{b})^{\frac{1}{\xi}-3}}{\left(a+\dfrac{b}{t}\right)^2}
\end{align}

\begin{align} \label{eq:31}
\Omega_{de}= \dfrac{2}{3} \dfrac{(m^2+4m+1)}{(m+1)^2}-\dfrac{\alpha^{2}(e^{at}t^{b})^{-2}}{\left(a+\dfrac{b}{t}\right)^2}-\dfrac{\rho_{0}(e^{at} t^{b})^{\frac{1}{\xi}-3}}{3 \left(a+\dfrac{b}{t}\right)^2}
\end{align}

and subsequently the total density parameter can be obtained as, 

\begin{align} \label{eq:32}
\Omega_{T} \equiv \Omega_{cs} + \Omega_{de} = \mathcal{A} +\dfrac{4m}{(m+1)^2} -\dfrac{\alpha^{2}(e^{at}t^{b})^{-2}}{\left(a+\dfrac{b}{t}\right)^2},
\end{align}

where, $\mathcal {A}$ = $\dfrac{2 \sigma^{2}}{3H^{2}}= \dfrac{2}{3} \left( \dfrac{m-1}{m+1} \right) ^{2}$ is the anisotropy parameter of expansion. Here also we can see that for $m=1$, the anisotropy parameter is zero, which further confirms the isotropic behaviour of the model.

\section{Physical Properties of the model}

\begin{figure}[h!]

\minipage{0.40\textwidth}
\includegraphics[width=75mm]{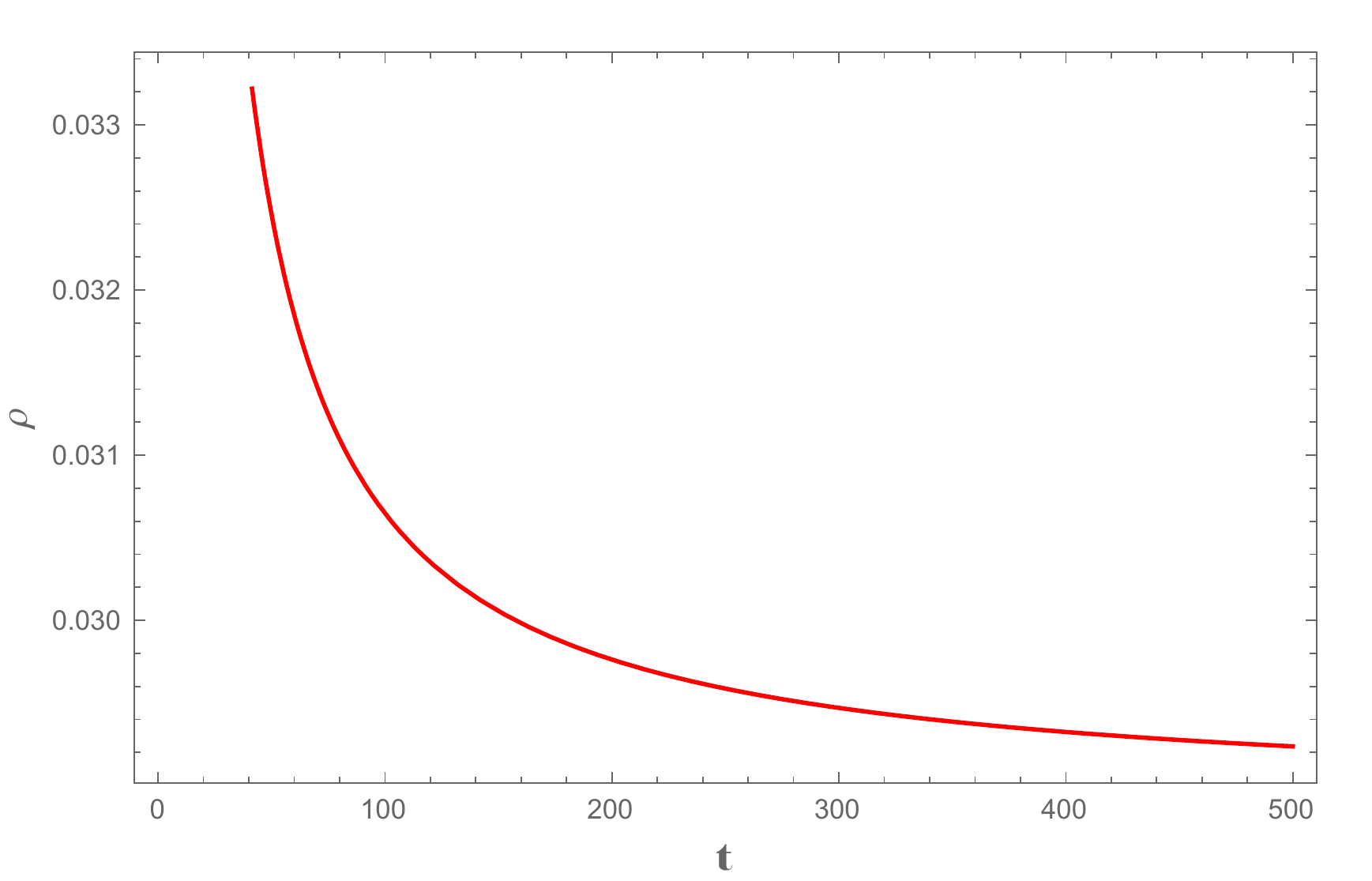}
\caption{Variation of DE density $\rho_{de}$ vs cosmic time with $m=2,$ $a=0.1,$ $b=0.3,$ $\alpha = 0.05,$ $\rho_{0}=0.2$ and $\xi =1$}
\endminipage
\end{figure}

From Fig. 2, we observed that during cosmic evolution, the DE density $(\rho_{de})$ remains positive. Therefore, in the derived model, both weak energy condition (WEC) and null energy condition (NEC) are satisfied. Further, $\rho_{de}$ decreases with time and leads to a small positive values in present time. The value of DE density comes closer to zero at late phase $(t \in [400,500])$ and then smoothly approaches to small positive value which indicates that the considered two fluid affects the DE density.  Similarly, it can be shown the string energy density $(\rho)$ will follow the same behaviour as of $\rho_{de}$. Fig. 3, depicts the variation of EoS parameters $(\omega_{de})$ with cosmic time, as a representative case with appropriate choice of constants of integration and other physical parameters for accelerating phase of universe.  The SN Ia data suggests the range value of EoS parameter as $-1.67 < \omega_{de} < -0.62$ where as the limit imposed by a combination of SN Ia data with CMB anisotropy and galaxy clustering statistics is in the range $-1.33 < \omega_{de} < -0.79.$ Again Fig. 3 shows that $\omega_{de}$ evolves in the range, which is in agreement with SN Ia and CMB observations. Also, the value of EoS parameter increases strictly in a range $(t \in [5,10])$, then decreases smoothly and remains constant in the preferred range. The behaviour of $\omega_{de}$ is different from other discussed literature but the value lies in the same range as in the observational result. So, we can infer here that the considered two fluid matter affects the EoS parameter of the model. \\

To check whether our model is realistic or not, the energy densities of the fluid need to be worked out. For a large value of $t$, the density parameter approaches to a finite value as $\Omega_{de}=\frac{2}{3}\frac{m^2+4m+1}{(m+1)^2}$. Further, the parameter started decreasing at an initial value and subsequently increases from the representing value of the cosmic time $t=0.462$ and remain constant at large $t$. In the case of string fluid, the density parameter $\Omega_{cs}$ initially increases and started decreasing at a representing value of $t=0.478$ and subsequently vanishes at infinite time. However,  the total density parameter is found to be $\Omega_T=0.97$, which also in accordance with the present observational data.  Also, it has been observed that the string density parameter dominates for a small time period i.e. at $t=0.471$ to $t=0.473$ and beyond that the DE density parameter dominates further.  The details are depicted in Fig. 4. \\

Fig. 5 shows that the evolution of skewness parameters with cosmic dynamics. In this figure, it can be noted that the behaviour of skewness parameters are totally controlled by the  parameter $m$. For $m=1$, the skewness parameters retain only the cosmic string nature giving an indication that the cosmic string matter affects the skewness parameters. We observed, at early cosmic phase $\eta$ starts with a small positive value, decreases with the cosmic time to became minimum and then becomes constant with further increase in cosmic time. The skewness parameter $\gamma$ also starts from a small negative value, attains maximum and becomes constant with respect to cosmic time. The evolutionary behaviour of $\delta$ is just the similar image of  $\delta$ but it retains a negative constant value at late phase. It can be noted that the pressure anisotropy factors along $x$, $y$ and $z$ axis $(\eta, \delta,\gamma)$ evolves with different nature, attain their extreme in a definite range of cosmic time $1<t<2$ and remain constant at later time. Hence we can infer that, the pressure anisotropies remain even at the late phase of cosmic evolution though in the early phase pressure was assumed to be isotropic. Also, it is interesting to note that, at the switching over cosmic time $t=0.02$, all the skewness parameters vanish.  

\begin{figure}[h!]
\minipage{0.40\textwidth}
\centering
\includegraphics[width=65mm]{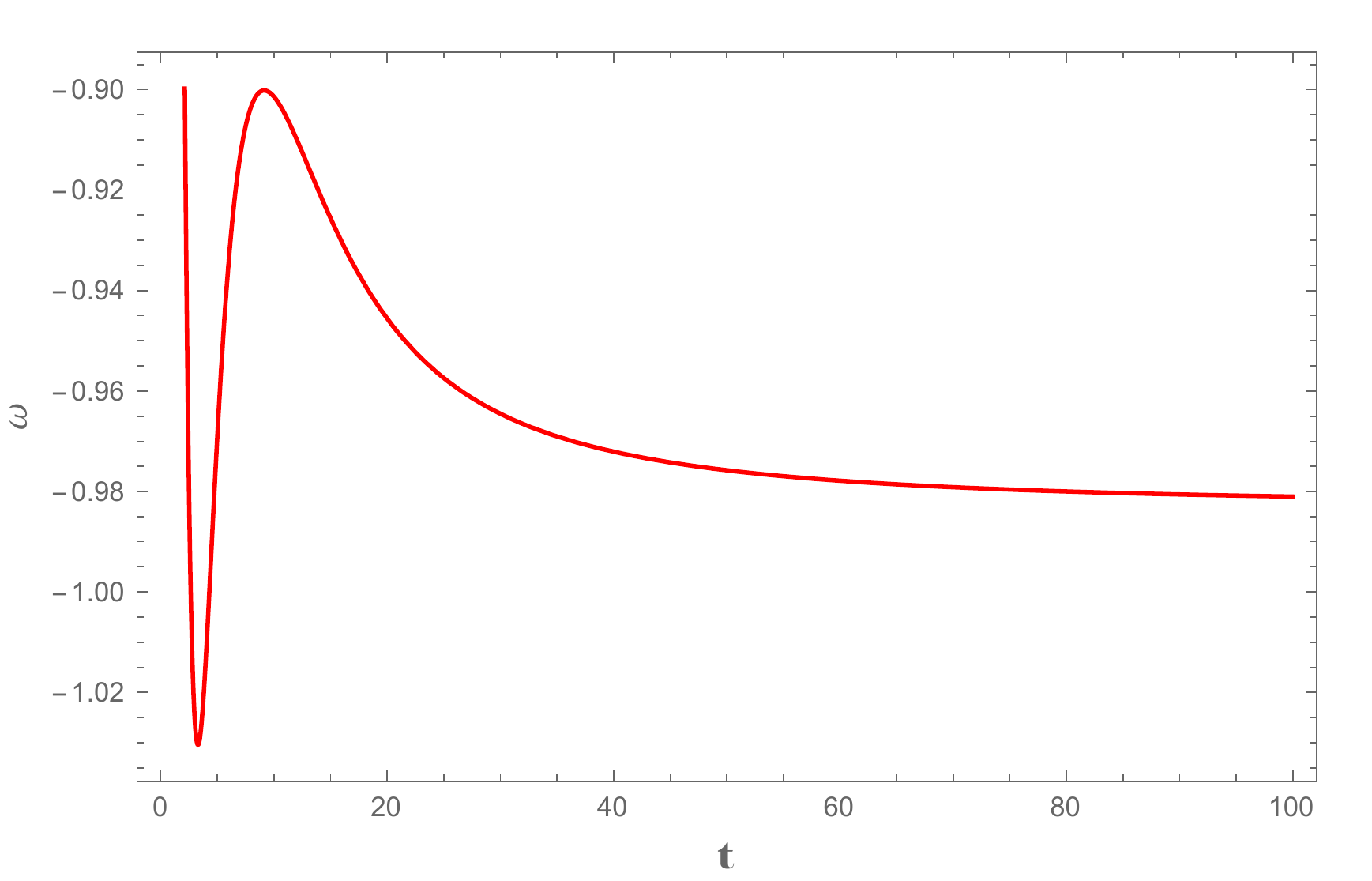}
\caption{Dynamics of EoS parameter of DE with respect to cosmic vs time with $m=2,$ $a=0.1,$ $b=0.3,$ $\alpha = 0.05,$ $\rho_{0}=0.2$ and $\xi = 1$}
\endminipage\hfill
\minipage{0.40\textwidth}
\includegraphics[width=65mm]{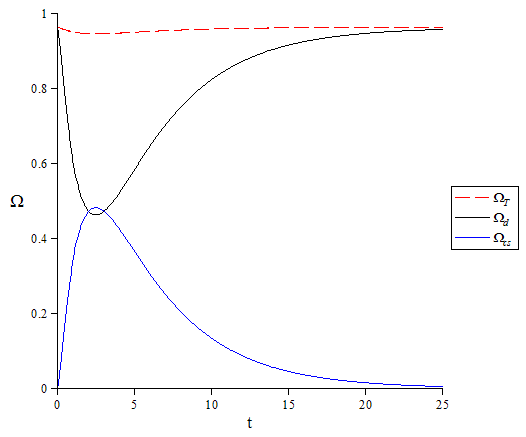}
\caption{Density parameter of DE $\Omega_{de}$, string fluid $\Omega_{cs}$ and the total density parameter $\Omega_T$ with respect to cosmic time with $m=2,$ $a=0.1,$ $b=0.3,$ $\alpha = 0.05,$ $\rho_{0}=0.2$ and $\xi = 1$}
\endminipage
\end{figure}

\begin{figure}[h!]
\minipage{0.40\textwidth}
\includegraphics[width=65mm]{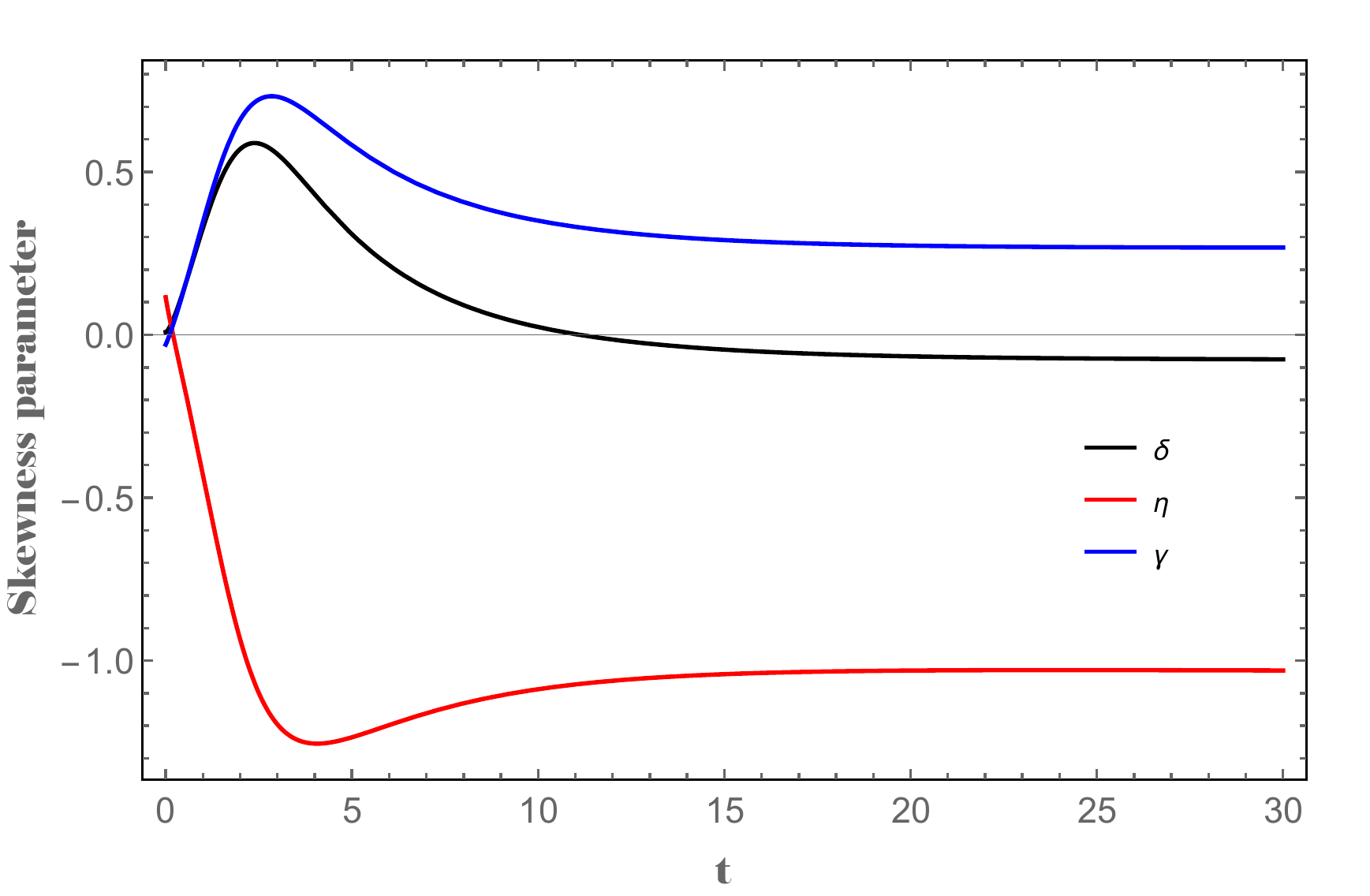}
\caption{Dynamical evolution of Skewness parameters along three spatial directions with respect to cosmic time with $m=2,$ $a=0.1,$ $b=0.3,$ $\alpha = 0.05,$ $\rho_{0}=0.2$ and $\xi = 1$}
\endminipage
\end{figure}

The scalar expansion $ \theta=3H=3(a+\frac{b}{t})$ and shear scalar $\sigma^2=\frac{1}{2}(H_x^2+H_y^2+H_z^2-\frac{\theta^2}{3})=\biggl(\frac{m-1}{m+1}\biggr)^2\biggl(a+\frac{b}{t}\biggr)^2 $  start off with extremely large values and continue to decrease with expansion of universe. The anisotropic parameter $\mathcal{A}_m=\frac{1}{3}\sum \biggl(\frac{\vartriangle H_i}{H}\biggr)^2=\frac{2}{3}\biggl(\frac{m-1}{m+1}\biggr)^2$, which is a measure of deviation from the isotropic expansion. If $\mathcal{A}=0$, the model is isotropic, otherwise anisotropic. In this case for $m=1$, the model becomes isotropic. Moreover, the average anisotropic parameter of the model is time independent which gives the indication that the anisotropy in expansion rates is maintained throughout the cosmic evolution. The state finder diagnostic pair $(r,s)$ provide an idea about the geometrical nature of the model and also through this the viability of the DE model can be tested. The pair can be obtained as $r=\frac{\tau^3-3b\tau+2b}{\tau^3}$ and $s= \frac{4b-6b\tau}{6b\tau-9\tau^3}$, where $\tau=at+b$. The state finder pair values depend on the constants $a$ and $b$ of the scale factor chosen. The state finder pair for the present model at the initial time is $\left({\frac{b^2-3b+2}{b^2}}, \frac{2}{3b}\right)$ and at the late time of cosmic evolution the value of state finder pair is $(1,0)$, which confirms that the model behaves like $\Lambda$ CDM.

\section{Conclusion}

In the present work, we have constructed an anisotropic DE cosmological model in the framework of GR and Bianchi type V space time. In order to study the evolution of skewness parameters and the EoS parameters, we have developed the DE models in two fluid situations. The EoS parameter of DE evolves within the range predicted by the observations. In the present epoch, the DE dominates the universe. This may be attributed to the current accelerated expansion of the universe. The parameter $m$ is responsible for the anisotropic behaviour and the matter behaviour of the model in the sense that, if $m=1$, we get isotropic model where only string matter part exist and for $m \neq 1,$ anisotropic and two fluid nature will be retained. In the present models,as expected, the matter energy density and DE density remains positive. Therefore, the energy conditions are satisfied, which in turn imply that the derived models are physically realistic. Moreover, it is showed that, there remain pressure anisotropies even at the late phase of cosmic evolution, though in the early phase pressure was assumed to be isotropic. Further the state finder pair is varying with time and decrease with the cosmic evolution in late time. Further the EoS of parameter changed with cosmic time and which indicates that during evolution there exists a relationship between the mean pressure of the cosmic fluid and energy density. The total density parameter $\Omega_T$, which quantifies the density of matter in the universe is found to be in the observational range $-1<\Omega_T<1$. In most part of the evolution, the DE density dominates the matter density; however for a small range of the cosmic time, the matter density dominates the DE density. So, we can conclude through the behaviour of the density parameter, that our model also mimics the cosmic acceleration in the early phase and also in late phase.\\


\begin{thebibliography}{99} 
\section*{References}

\bibitem{riess1} A. G. Riess, A. V. Filippenko, P. Challis et al., \textit{The Astronomical Journal}, \textbf{116}(3), 1009, 1998.

\bibitem{perl1} S. Perlmutter, G. Aldering, G. Goldhaber et al.,\textit{ The
Astrophysical Journal}, \textbf{517},(2), 565, 1999.

\bibitem{Soleh66} J.E. Solheim, \textit{Mon. Not. R. Astron. Soc.}, \textbf{133}, 321, 1966.

\bibitem{perl2} S. Perlmutter et al., \textit{The Astrophysics Journal}, \textbf{483}, 565, 1997.

\bibitem{perl3} S. Perlmutter et al., \textit{The Nature}, \textbf{391}, 51, 1998.

\bibitem{riess2} A. G. Riess et al., \textit{The Astronomical Journal}, \textbf{607}, 665, 2004.

\bibitem{cald1} R. R. Caldwell and M. Doran, \textit{Physics Review D}, \textbf{69}, 103517, 1998.

\bibitem{huang} Z. Y. Huang, B. Wang et al., \textit{Journal of Cosmology and Astroparticle Physics}, \textbf{05}, 013, 2006.

\bibitem{seljak} U. Seljak et al., \textit{Physics Review D}, \textbf{71}, 103515, 2005.

\bibitem{tegmark} M. Tegmark et al. (SDSS Collaboration), \textit{Physics Review D}, \textbf{69}, 103501, 2004.

\bibitem{eisenstein} D. J. Eisenstein et al. (SDSS Collaboration), \textit{The Astronomical Journal}, \textbf{633}, 560, 2005.

\bibitem{jain} B. Jain and A. Taylor, \textit{Physics Review Letter}, \textbf{91}, 141302, 2003.

\bibitem{astier1} P. Astier and R. Pain, \textit{1204.5493[astr-ph.CO].}, 2012.

\bibitem{copeland} E. J. Copeland, M. Sami and S. Tsujikawa \textit{International Journal of Modern Physics D}, \textbf{15}, 1753, 2006.

\bibitem{cald2} R. R. Caldwell and M. Kamionkowski, \textit{Annual Review of Nuclear and Particle Science}, \textbf{59}, 397, 2009.

\bibitem{silve} A. Silvestri, M. Trodden, \textit{Reports On Progress in Physics}, \textbf {72}, 96901, 2009.

\bibitem{ratra} B. Ratra, P. J. E. Peebles, \textit{Physics Review D}, \textbf{37}, 321, 1998.

\bibitem{sahni} V. Sahni, A. Starobinsky \textit{International Journal of Modern Physics D}, \textbf{9}, 373, 2000.

\bibitem{cald3} R. R. Caldwell, \textit{Physics Letter B}, \textbf{545}, 23, 2002.

\bibitem{armen1} C. Armendariz-Picon, V. Mukhanov and P. J. Steinhardt, \textit{Physics Review Letter}, \textbf{85}, 4438, 2000.

\bibitem{armen2} C. Armendariz-Picon, V. Mukhanov and P. J. Steinhardt, \textit{Physics Review Letter}, \textbf{63}, 103510, 2001.

\bibitem{sen} A. Sen, \textit{Journal of High energy Physics}, \textbf{204}, 48, 2002.

\bibitem{feng} B. Feng, X.L. Wang and X.M. Zhang, \textit{Physics Letter B}, \textbf{607}, 35, 2005.

\bibitem{bento} M. C. Bento, O. Bertolami and A. A. Sen, \textit{Physics Review D}, \textbf{66}, 43507, 2002.

\bibitem{kamen} A. Y. kamenshchik, M. Moschella and V. Pasquier, \textit{Physics Letter B}, \textbf{511}, 265, 2001.

\bibitem{wang} B. Wang, Y. G. Gong and E. Abdalla, \textit{Physics Letter B}, \textbf{624}, 141, 2005.

\bibitem{setare1} M. R. Setare, \textit{Physics Letter B}, \textbf{642}, 421, 2006.

\bibitem{setare2} M. R. Setare, \textit{Physics Letter B}, \textbf{644}, 99, 2007.

\bibitem{daff} C. Daffayet, G. R. Dvali, G. Gabadadze, \textit{Physics Review D}, \textbf{65}, 44023, 2002.

\bibitem{li} M. Li, \textit{Physics Letter B}, \textbf{603}, 1, 2004.

\bibitem{astier2} P. Astier et al., \textit{Astronomy and Astrophysics}, \textbf{447}, 31, 2006.

\bibitem{mac} C. J. MacTavish et al., \textit{The Astrophysical Journal}, \textbf{647}, 799, 2006.

\bibitem{koma} E. Komatsu et al., \textit{The Astrophysical Journal Supplement Series}, \textbf{180}, 330, 2009.

\bibitem{knop} R. K. Knop et al., \textit{The Astrophysical Journal}, \textbf{598}, 102, 2003.

\bibitem{campa1} L. Campanelli, P. Cea and L. Tedesco, \textit{Physics Review Letter}, \textbf{97}, 131302, 2006.

\bibitem{campa2} L. Campanelli, P. Cea and L. Tedesco, \textit{Physics Review D}, \textbf{76}, 63007, 2007.

\bibitem{campa3} L. Campanelli, \textit{Physics Review D}, \textbf{80}, 63006, 2009.

\bibitem{gruppo} A. Gruppo, \textit{Physics Review D}, \textbf{76}, 83010, 2007.

\bibitem{mishra1} B. Mishra, P. K. Sahoo and S. K. Tripathy, \textit{Astrophysics Space Science}, \textbf{356}, 163, 2015.

\bibitem{pacif} S. K. J. Pacif, B.Mishra \textit{Astrophysics Space Science}, \textbf{360}, 48, 2015.

\bibitem{mishra2} B. Mishra, S. K. Tripathy, \textit{Modern Physics Letter A}, \textbf{30}, 1550175, 2015.

\bibitem{letelier} P. S. Letelier, \textit{Physics Review D}, \textbf{28}, 2414, 1983.

\bibitem{stachel} J. Stachel, \textit{Physics Review D}, \textbf{21}, 217, 1980.

\bibitem{akarsu1} O. Akarsu and C. B. Kilinc, \textit{GR and Gravitation}, \textbf{42}, 119, 2010.

\bibitem{akarsu2} O. Akarsu and C. B. Kilinc, \textit{General Relativity and Gravitation}, \textbf{42}, 763, 2010.

\bibitem{yadav1} A. K. Yadav, F. Rahaman and S. Ray, \textit{International Journal of Theoretical Physics}, \textbf{50}, 871, 2011.

\bibitem{amir1} H. Amirhashchi, \textit{Astrophysics Space Science}, \textbf{345}, 439, 2013.

\bibitem{amir2} H. Amirhashchi, A. Pradahan and H. Zainuddin, \textit{Research in Astronomy and Astrophysics}, \textbf{13}, 119, 2013.

\bibitem{pradhan1} A. Pradhan, H. Amirhashchi and B. Saha, \textit{Astrophysics Space Science}, \textbf{333}, 343, 2011.

\bibitem{pradhan2} A. Pradhan, H. Amirhashchi and B. Saha, \textit{International Journal of Theoretical Physics}, \textbf{50}, 2923, 2011.

\bibitem{shey} A. Sheykhi and M. R. Setare, \textit{Modern Physics Letter A}, \textbf{26}, 1897, 2011.

\bibitem{zimda} W. Zimdahl and D. Pavon, \textit{General Relativity and Gravitation}, \textbf{36}, 1483, 2004.

\bibitem{tripa1} S. K. Tripathy, B. Mishra and P.K. Sahoo, \textit{arXiv:1503.05109v1 [General Relativity-qc]}, 2015.

\bibitem{tripa2} S. K. Tripathy, \textit{Astrophysics Space Science}, \textbf{350}, 367, 2014.

\bibitem{akarsu3} O. Akarsu, S. Kumar, R. Myrzakulov, M. Sami and L. Xu, \textit{Journal of Cosmology and Astroparticle Physics}, \textbf{01}, 22, 2014.

\bibitem{kumar2} S. Kumar, \textit{Gravitation and Cosmology}, \textbf{19}, 284, 2013.

\bibitem{pradhan3} A. Pradhan, A.K. Pandey, R.K. Mishra, \textit{Indian Journal of Physics}, \textbf{350}, 367, 2014.

\bibitem{rapet} D. Rapetti, S.W. Allen, M.A. Amin, R.D. Blanford, \textit{Monthly Notices of the Royal Astronomical Society}, \textbf{375}, 1510, 2007. 

\bibitem{giost}R. Giostri et al., \textit{Journal of Cosmology and Astroparticle Physics},\textbf{1203}, 027, 2012.

\end{thebibliography}
\end{document}